\documentclass[12pt]{article}
\usepackage{aasms}
\usepackage[dvips]{graphicx}
\usepackage{rotating}
\usepackage{lscape}
\usepackage{doublespace}

\begin{document}

\singlespace

\title{Nature vs. Nurture: The Origin of Soft Gamma--ray 
Repeaters and Anomalous X--ray Pulsars}

\author{D. Marsden\footnote[1]{NAS/NRC Research Associate}}
\affil{NASA/Goddard Space Flight Center, Code 662, Greenbelt, MD 20771}

\author{R. E. Lingenfelter and R. E. Rothschild}
\affil{Center for Astrophysics and Space Sciences, University of 
California at San Diego \\ La Jolla, CA 92093}

\author{J. C. Higdon}
\affil{W. M. Keck Science Center, Claremont Colleges, Claremont, 
CA 91711}

\begin{abstract}

Soft gamma--ray repeaters (SGRs) and anomalous x--ray pulsars 
(AXPs) are young and radio-quiet x-ray pulsars which have been 
rapidly spun-down to slow spin periods clustered in the range 
$5-12$ s. Most of these unusual pulsars also appear to be associated 
with supernova shell remnants (SNRs) with typical ages $<30$ kyr. 
By examining the sizes of these remnants versus their ages, we 
demonstrate that the interstellar media which surrounded the SGR 
and AXP progenitors and their SNRs were unusually dense compared 
to the environments around most young radio pulsars and SNRs. 
We explore the implications of this evidence on magnetar and 
propeller-based models for the rapid spin-down of SGRs and AXPs. 
We find that evidence of dense environments is not consistent with 
the magnetar model unless a causal link can be shown between the 
development of magnetars and the external ISM. Propeller-driven 
spin-down by fossil accretion disks for SGRs and AXPs appears to 
be consistent with dense environments since the environment can 
facilitate the formation of such a disk. This may occur in two 
ways: 1) formation of a ``pushback'' disks from the innermost 
ejecta pushed back by prompt reverse shocks from supernova 
remnant interactions with massive progenitor wind material 
stalled in dense surrounding gas, or 2) acquisition of disks 
by a high velocity neutron stars, which may be able to capture 
a sufficient amounts of co-moving outflowing ejecta slowed by 
the prompt reverse shocks in dense environments.  

\end{abstract}


\section{Introduction}
\label{intro}

Soft gamma--ray repeaters (SGRs) are neutron stars whose multiple 
bursts of gamma--rays distinguish them from other gamma--ray burst 
sources (e.g \cite{hurley2000a} for a recent review). SGRs are also 
unusual x--ray pulsars in that they have spin periods clustered in the 
interval $5-8$ s, and they all appear to be associated (\cite{cline82}; 
\cite{felten81}; \cite{kulkarni93}; \cite{vasisht94}; \cite{hurley99a}; 
\cite{hurley99c}; \cite{corbel99}; \cite{cline00}) with supernova 
remnants (SNRs), which limits their average age to approximately 
$20$ kyr (\cite{braun89}). The angular offsets of the SGRs from 
the apparent centers of their associated supernova remnant shells 
indicate that SGRs are endowed with space velocities $>500$ km 
s$^{-1}$, which are greater than the space velocities of most 
radio pulsars (\cite{cordes98}). Anomalous x--ray pulsars (AXPs) 
are similar to SGRs in that they are radio quiet x--ray pulsars 
with spin periods clustered in the range $6-12$ s, and have similar 
persistent x--ray luminosities as the SGRs ($\sim 10^{35}$ ergs 
s$^{-1}$, see e.g. \cite{stella98} for a recent review). Most of 
the AXPs appear to be associated with supernova remnants, and 
therefore they are also thought to be young neutron stars like 
the SGRs. The spin periods of both AXPs and SGRs are increasing with 
time (spinning-down), and show no evidence for intervals of decreasing 
spin period (spin-up), although the spin-down rates of many of the 
SGRs and AXPs appear to be variable or ``bumpy'' (e.g. \cite{baykal96}; 
\cite{marsden99}; \cite{woods99c}; \cite{woods00}; although for a 
different viewpoint see \cite{kaspi99}).

The lack of identified companions at non x--ray wavelengths (e.g. 
\cite{stella98}) and Doppler shifts associated with binary orbital 
motion (\cite{mer98}), together with the problem of accelerating 
binaries to space velocities $> 1000$ km s$^{-1}$ (\cite{rothschild94}), 
imply that SGRs and AXPs are not members of high mass binary systems, 
although low mass systems with neutron star companions of $<1M_{\odot}$ 
are not constrained in most cases. If SGRs and AXPs spin-down primarily 
via the emission of magnetic dipole radiation (MDR), as do radio pulsars, 
then they must have surface dipole fields of $\sim 10^{14}$ G or greater 
(i.e. ``magnetars''; e.g. \cite{thompson95}). Observations of the SGRs 
1806--20 (\cite{kouv98}) and 1900+14 (\cite{marsden99}; \cite{woods99c}), 
however, indicate that the present-day spin-down rates of these 
SGRs are inconsistent with simple MDR, given the ages of their 
associated supernova remnants (\cite{harding99}; \cite{rothschild99}), 
and imply that the spin-downs are due to winds. Magnetar-strength 
fields might still be possible in these sources, however, if Alfv\'{e}n 
wave wind emission is infrequent and intermittent (\cite{harding99}, so 
that the presently observed spin-down rates are atypical. Alternative 
scenarios for SGRs and AXPs involving typical pulsar magnetic fields 
($\sim 10^{12}$ G) have been proposed (\cite{vanparadijs95}; Alpar 2000; 
Chatterjee, Hernquist \& Narayan 2000; Chatterjee \& Hernquist 2000). 
In these models, the SGRs and AXPs have spun-down rapidly via 
magnetospheric accretion torques from outflowing ``propeller effect'' 
winds. The assumed sources of the magnetospheric  material are either 
fallback accretion disks (Alpar 2000; Chatterjee, Hernquist \& Narayan 
2000; Chatterjee \& Hernquist 2000), or fossil disks formed from 
expanding supernova ejecta intercepted by high velocity neutron stars 
(\cite{vanparadijs95}; \cite{corbet95}). 

Here we present a fresh look at evidence which suggests that SGRs and 
AXPs are born into unusually dense environments. We show that the 
environments of the SGR and AXP progenitors into which their SNRs 
expand are the dense, warm and cool phases of the interstellar medium 
(ISM), and not the hot tenuous phase of the ISM where most of 
the neutron-star-producing, core collapse supernovae of massive O and B 
stars occur and where most young radio pulsars are found. This implies 
that there is an environmental factor influencing the development of 
SGRs and AXPs. The structure of this paper is as follows. We first 
discuss the typical environments of supernova progenitors in $\S 2$, 
and then supernova remnants associated with SGRs and AXPs in $\S 3$, 
followed by a discussion of the SGR and AXP ages and distances are in 
$\S 4$. In $\S 5$ the density of the SGR and AXP progenitor environments 
is discussed, and in $\S 6$ a similar analysis is done for the SNRs 
associated with young radio pulsars. In $\S 7$ we discuss the 
statistical significance of the results and the implications for 
magnetar and propeller-based models for SGRs and AXPs. Finally, the 
Appendix contains a short discussion of the ages, distances, and 
other information for each SGR and AXP. 

\section{The Environments of Supernova Progenitors}
\label{environment}

What are the environments typical of neutron star progenitors? 
Observations clearly show that the majority of neutron stars are 
formed in ``superbubbles'': evacuated regions of the ISM which 
surround the OB associations in which the massive progenitors of 
most neutron stars live. This is because most O and B stars ($>$ 
80\%) are observed (e.g. McCray \& Snow 1979) to occur in clusters 
formed from giant molecular clouds ($> 10^5 M_{\odot}$); much smaller 
clouds are disrupted by the radiation and winds from the first O star 
that forms. These massive ($>8$ M$_{\odot}$: \cite{woosley95}) and 
slow-moving ($\sim 4$ km s$^{-1}$: \cite{blaauw61}) O and B star 
progenitors of Type II and Ib/c supernovae do not travel far 
from their birthplaces during their relatively short ($<30$ Myr: 
\cite{schaller92}) lives. The supernovae from these massive stars 
are therefore heavily clustered in space and time and form vast 
($>100$ pc) HII regions/superbubbles (e,g. \cite{maclow88}) filled 
with a hot ($>10^{6}$ K) and tenuous ($n\sim 10^{-3}$ cm$^{-3}$) 
gas. 

There is also direct observational evidence that the great majority 
($>$80\%) of neutron star stars are born into superbubbles 
consisting of hot and diffuse ISM gas. Observations of a sample 
of 49 spectroscopically identified Type II and Ib/c, core-collapse 
supernovae occurring in face-on late-type spiral galaxies by van 
Dyk et al. (1996) found that $72\%\pm 10\%$ of Type II and $68\%\pm 
12\%$ of Type Ib/c supernovae are in resolvable giant HII region 
superbubbles. We suggest that these fractions are, in fact, only 
a lower limit on the occurrence of core-collapse supernovae in 
superbubbles, because of the difficulty in detecting faint HII 
regions in distant galaxies. Correcting for this effect using 
the H$\alpha$ luminosity distributions of Kennicutt, Edgar, \& 
Hodge (1989), we find that the H$\alpha$ threshold used by van 
Dyk et al. (1996) would have allowed them to resolve only $76\%$ 
of the HII regions in their sample galaxies. This clearly suggests 
that the great majority -- i.e. $90\%\pm 10\%$ -- of neutron star 
progenitors reside in the hot and diffuse ($n<0.01$ cm$^{-3}$) 
ISM in superbubbles, with $<20\%$ occurring in the denser phases 
of the ISM.

When one corrects for selection effects, the distribution of surface 
brightness of all detected Galactic supernova remnants provides another 
measure of the fractions of supernovae occurring in the warm and hot 
phases of the ISM. Detections of radio remnants are limited by surface 
brightness, which is independent of their distance, and SNRs in the 
denser ISM have larger surface brightnesses than SNRs of the same age 
in the hot/diffuse ISM (e.g. \cite{gull73}). A study (\cite{higdon80}) 
of the age versus surface brightness of the remnants of historical 
supernovae, using Gull's model, suggests that the maximum detectable 
ages of radio supernova remnants above a nominal surface brightness 
density of 10$^{-21}$ W m$^{-2}$ Hz$^{-1}$ sr$^{-1}$ at 408 MHz 
are roughly 20 kyr in the warm ($n=0.1$ cm$^{-3}$) gas and 6 kyr 
in the hot superbubble ($n=0.001$ cm$^{-3}$) gas. Combining this 
with an analysis of the number distribution of SNRs as a function 
of surface brightness suggests that about $30\%$ of the detected 
radio SNRs are in the warm denser ($n>0.1$ cm$^{-3}$) phase of the 
ISM (e.g. \cite{higdon80}). This implies that only about $10\%$ 
of the supernovae occur in the warm ($n>0.1$ cm$^{-3}$) ISM
\footnote[2]{We should mention that Higdon \& Lingenfelter (1980) 
further suggested that this fraction represented the filling factor 
of the warm ISM, on the mistaken assumption that the supernovae 
were uniformly distributed, rather than being predominantly 
clumped in OB associations.}, and about $90\%$ in the hot phase, 
since, for a total (including Type Ia) SN rate of 1 SN every 
30 yrs, there are then roughly 180 SNR younger than 6 kyr that 
are detectable in the hot superbubble ISM, and roughly 70 SNR 
younger than 20 kyr that are detectable in the denser warm ISM, 
or about $30\%$ of the detected radio SNRs in the warm ISM -- 
as the observed distribution implies.

\section{The Supernova Remnants Associated With SGRs and AXPs}
\label{snrs}

The environments of SGRs and AXPs are probed by the blastwaves of 
their associated supernova remnants, and from the size of the remnant 
shell as a function of the age we can constrain the external density. 
In Table $1$ we have listed the $12$ known SGRs and AXPs and their 
associated supernova remnant shells. For more information on the 
individual SGRs and AXPs, see the recent reviews of Hurley (2000a) 
and Mereghetti (1999), and the Appendix. We include the source AXP 
1845--0258 -- although no period derivative has been measured for this 
source -- because it is commonly cited as an AXP (see e.g. the 
discussion in \cite{gaensler99}). We also include the AXP 0720--3125 
for completeness, although its identification as an AXP has been 
questioned as well (see Appendix). The new and tentative SGR 1801--23 
has also been included on the basis of two soft gamma--ray bursts 
consistent with a single location on the sky (\cite{cline00}).

The identification of the associated remnants are based on positional 
coincidences between the remnant and the SGR/AXP, and in some cases 
on similar distances as implied by the hydrogen column densities measured 
from the x--ray spectrum of the SGR/AXP and its associated remnant. 
Based on these criteria, associated SNRs have been previously 
been identified for 7 out of the 12 SGRs and AXPs (see Table 1 and 
Appendix). Using these same criteria, we suggest three additional 
SNR associations with SGR/AXPs (also included in Table 1 and Appendix). 
We suggest two new AXP/SNR associations, AXP 1709--40 with the SNR 
G346.6--0.2 (\cite{green98} and references therein) and AXP 1048--5937 
with G287.8--0.5 (\cite{jones73}), based on the near coincident 
positions between each AXP and a known supernova remnant, and also 
(for AXP 1048--5937/G287.8--0.5) on similar implied distances for the 
AXP and the remnant. These two associations may have been discounted 
previously because they would imply larger than average neutron star 
velocities, but in view of the comparable velocities implied for three 
of the SGRs these possible associations should be considered. AXP 
1048--5937 lies $2.2$ shell radii from the apparent center of the 
SNR G287.8--0.5, which is associated with the Carina Nebula, a region 
of massive star formation at a distance $\sim 2.5-2.8$ pc (\cite{the71}; 
\cite{seward82}). This relatively small distance of the SNR is quite 
consistent (see Fig. 2) with the low $N_{H}=(0.45\pm 0.10)
\times 10^{22}$ cm$^{-2}$ found from the AXP 1048--5937 x--ray spectral 
fits (\cite{oos98}). For an estimated SNR age of $\sim$20 kyr, the 
displacement of the AXP implies a relatively modest neutron star 
velocity of $\sim 1000$ km s$^{-1}$. A similar transverse velocity 
is indicated by the association of AXP 1709--40, which lies 1.7 shell 
radii from the center of the well defined shell remnant G346.6--0.2, 
assuming a similar age and a distance of 3 to 5 kpc, based on 
Galactic structure. Such a SNR distance is quite consistent with 
its association with AXP 1709--40, which has an x-ray absorption 
$N_{H}=(1.81\pm 0.07)\times 10^{22}$ cm$^{-2}$ (\cite{sug97}). We 
also propose that the new SGR candidate 1801--23 (\cite{cline00}) may 
be associated with the well studied SNR W28. Although there is no 
distance estimate to the SGR (\cite{cline00}), W28 is the only known 
SNR through which the thin SGR error box passes. Therefore we regard 
the tentative W28/SGR 1801--23 association as promising and encourage 
observations to search for x--ray point sources (e.g. \cite{andrews83}) 
within the remnant which may be the SGR counterpart. We include these 
three tentative SNR associations because they are suggestive, but 
as we show in $\S 7.1$ these data can be left out of the sample 
without affecting the overall conclusions of our analysis.

Using the SNR catalog of Green (2000), we are unable to identify 
any possible remnants associated with AXPs 0720--3125 and 0142$+$615, 
which may be the two oldest AXPs based on their pulsar timing ages 
$0.5P/\dot{P}$, where $P$ is the pulsar period and $\dot{P}$ is the 
period derivative, of $40$ kyr (\cite{haberl97}) and $60$ kyr (e.g. 
\cite{white96}), respectively. For AXP 0720--3125 the lack of an 
associated remnant is not surprising, because this source is thought 
to be so close ($\sim 100$ pc) that a neutron star with a velocity 
of $\sim$ 1000 km s$^{-1}$ would have traveled $\sim 40$ pc in $40$ 
kyr and hence could have originated from a large area of the sky. In 
addition, the large scale radio surveys of the Galactic Plane appear 
to be incomplete for very large ($>1^{\circ}$) diameter remnants (e.g. 
\cite{duncan97}), indicating that such a remnant could easily go 
undetected. AXP 0142$+$615 is situated in or behind a molecular cloud 
complex (\cite{israel94}; \cite{white96}), and if its associated 
remnant expanded in high density material it may have already passed 
into the radiative phase and faded below the surface brightness 
detection threshold. Therefore the detection of the remnants 
associated with both AXP 0720--3125 and AXP 0142$+$615 would 
be difficult, and we can not assign meaningful limits to the 
physical size of their (unknown) associated remnants, given the 
present data. We encourage new deep observations of the regions 
surrounding these objects to look for associated supernova remnants. 

We see from Table 1 that 10 out of the 12 SGR/AXPs have proposed 
associations with radio shell SNRs. The probabilities of a chance 
coincidence for some of the individual associations have been 
estimated (e.g. \cite{cline82}; \cite{felten81}; \cite{kulkarni93}) 
at a few times 10$^{-2}$ or less (see Appendix). The chance probability 
for an association between an SGR/AXP and an SNR can be estimated by 
simply considering the spatial density of known supernova remnants in 
the Galactic plane (in the following we assume that the localization 
error box of the SGR/AXP is small compared to the size of the SNR). 
For the extreme case, we consider just the SGRs and AXPs located in the 
highly crowded inner Galaxy. There all $8$ SGR/AXPs appear to have 
associated SNRs. These have Galactic longitudes between 286$^{\circ}$ 
and 43$^{\circ}$ and latitudes within $\pm$ 1.2$^{\circ}$. Within this 
area of 1.0$\times$10$^6$ arc min$^2$, there are 142 known SNRs (Green 
2000) covering a combined surface area of 3.4$\times$10$^4$ arc 
min$^2$, which is actually an overestimation due to overlap between 
the remnants. For a single association with the SGR or AXP displaced 
from the center of the remnant by as much as $2.3$ times the remnant 
radius (as implied by the SGR 1627--41/G337.0--0.1 association), the 
chance association probability is roughly $2.3^2\times 3.4\times 
10^{-2}\sim 18\%$, or $\sim 16\%$ correcting for overlap. Thus there 
is a significant probability that a single association might be simply 
a chance coincidence, if we relied solely on position, but the 
probability of getting even 4 chance associations out of 8 tries 
is only 2\%, and the probability that 8 out of 8 are spurious is 
just $4\times10^{-7}$.

\section{The Distances and Ages of the Associated Remnants}
\label{distances}

In order to probe the SNR environments, we need to know both the 
radii of the associated SNR shells, which can be determined from their 
measured angular diameters, and their estimated distances. These 
distances are listed in Table 2, and discussed in the Appendix. 
All but three of the distances were determined from Galactic kinematic 
arguments based on interactions or associations with adjacent objects 
with known distances (e.g. HII regions or molecular clouds), or on 
absorption line measurements. We did not use distances estimated
from the standard SNR surface brightness/diameter relationships,
because we did not want to bias the SNR sizes toward SNRs in the 
dense ISM, which is where most of the observed radio SNRs are 
located (e.g. Higdon \& Lingenfelter 1980; Kafatos et al. 1980).
SNRs expanding in the tenuous hot ISM become large and hard to detect 
(e.g. \cite{duncan97}) much quicker than SNRs in dense environments
and are under-represented. 

The positions of the Galactic SNRs associated with SGRs and AXPs 
are shown in Figure $1$, projected onto a model of the Galactic 
plane distribution of free electrons (\cite{taylor93a}) which 
traces star formation regions delineating the spiral arms (e.g. 
Georgelin \& Georgelin 1976). We see that the estimated distances 
of the SNRs are quite consistent Galactic structure, placing them 
in or near the spiral arms where the bulk of recent massive star 
formation has occurred. We also see that the estimated distances 
of the SNRs are not systematically underestimated in our analysis, 
since 5 out of the 9 SNR-AXP/SGR associations lie roughly at or 
beyond the distance of the Galactic center -- as would be expected 
for a relatively unbiased sample of Galactic sources. This also 
shows that they are not likely to be systematically much farther 
away, or most would lie on the other side of the Galaxy or even 
outside of the Galaxy.

To further test both the AXP/SGR--SNR associations and their estimated
distances, we compare the electron column depth associated with the
SNR distance with the H column depth determined from the x-ray absorption
of the AXP/SGR emission. We calculate the dispersion measure to each 
source given the assumed distances and compare that to the measured 
$N_{H}$ value for each source. Table $2$ lists the distance ranges, 
calculated dispersion measures using the Taylor \& Cordes (1993) 
model, and observed $N_{H}$ values for each SGR/AXP and associated SNR. 
The dispersion measure (DM) vs. $N_{H}$ is plotted in Figure $2$. As 
seen from the Figure, the data points are well correlated between 
lines representing free electron to hydrogen ratios $e/H=0.03-0.07$, 
which is consistent with the range of $0.02-0.08$, expected from the 
mean mass fraction of ionized gas in the Galaxy of $\sim 1-5\%$ (e.g. 
\cite{spitzer78}). Thus we see that each of associations is consistent 
both in proximity of direction and in similarity of distance. This 
also shows that the estimated distances to the SGRs/AXPs and their 
associated remnants are not systematically underestimated. More 
information on the $N_{H}$ values for each source is given in the 
Appendix. 

The ages of the most of the SGRs and AXPs are much more uncertain than 
their distances. This is because the usual pulsar timing age formula 
-- appropriate for pulsar spin-down due to magnetic dipole radiation 
(MDR) torque (although \cite{gaensler00}) -- does not seem to be 
applicable for SGRs and AXPs, in general, because it yields age 
estimates which are inconsistent with the likely ages of the 
associated supernova remnants and the implied neutron star 
velocities for at least three of the sources (SGR 1806--20, 
SGR 1900+14, and AXP 2259+58.6: \cite{kouv98}; \cite{marsden99}; and 
\cite{corbet95}, respectively). In addition, all of the SGR and AXPs 
for which pulsar braking indices have been measured  -- AXPs 1709--40 
and 2259+58.6 (\cite{kaspi99}) and SGR 1806--20 (\cite{woods00}) -- 
have braking indices which are inconsistent with simple MDR spin-down. 
The MDR timing age for AXP 1841--045 of 4000 yr, however, is consistent 
with the estimated age of the associated SNR Kes 73 (\cite{gotthelf99}), 
but the braking index for this AXP has not yet been measured. We 
therefore adopt ages for the SGRs and AXPs based on the ages of 
their associated SNRs. Some of the associated remnants (N49, Kes 
73, G29.6$+$0.1, and CTB 109) are relatively well-studied and have 
age estimates based on observed shock velocities and/or x--ray 
temperatures. For these remnants we take the minimum and maximum 
values for the ages of each remnant from the literature. For the 
rest of associated SNRs, we adopt a lower age limit of $t_{low}=
\rm{min}[t_{fe},t_{v}]$, where $t_{fe}=d_{min}\theta_{SNR}/v_{ej}$ 
is the age of the remnant assuming free expansion, with $d_{min}$ the 
minimum estimated distance, $\theta_{SNR}$ the SNR radius in radians, 
and $v_{ej}\sim 10^{4}$ km s$^{-1}$ is the maximum ejecta speed in free 
expansion. In addition, if we assume that the transverse velocity of 
the associated SGR or AXP cannot exceed $v_{max}$, the minimum age 
is $t_{v}=d_{min}\theta_{\ast}/v_{max}$, where $\theta_{\ast}$ is 
angular offset of the pulsar from the center of the remnant. In this 
paper we will assume $v_{max}=2000$ km s$^{-1}$, which exceeds the 
observed velocity of any pulsar (e.g. \cite{cordes98}). For the upper 
age limit of the SNRs we choose $30$ kyr, which is the maximum estimated 
age for the SNR/radio pulsar associations shown in Table 3\footnote[3]{The 
overall conclusions are unaffected if we instead choose $20$ kyr 
(\cite{braun89}) for the maximum observable SNR age, as discussed 
in $\S$ \ref{significance}}. For each SNR, we assume that the most 
likely age is the arithmetic mean of the age range, which is equivalent 
to assuming that the supernova rate is uniform in time throughout the 
Galaxy. 

\section{The Densities of SGR/AXP Progenitor Environments}
\label{densities} 

In Figure $3$ we have plotted the SNR shell radii $R_{SNR}$ as a 
function of the estimated age $t$ of each remnant associated with 
an SGR or AXP. Overplotted with solid lines are simple approximations 
of the evolutionary tracks (\cite{shull89}) of supernova remnant 
expansion in a wide range of the external ISM densities $n$, 
assuming a total supernova kinetic energy of $10^{51}$ ergs. These 
SNR evolutionary tracks move through three phases: the initial free 
expansion/ejecta-dominated phase of the remnant, where the mass of 
the SN ejecta is much greater than the mass of the swept-up ISM and 
$R_{SNR}\propto t$; the Sedov/adiabatic phase, which begins when the 
mass of the swept-up ISM is roughly $>10\%$ of the mass of the SN 
ejecta, and the remnant slows down with $R_{SNR}\propto t^{2/5}$; 
and finally the radiative/snowplow phase, where the shell of swept-up 
ISM radiates away the energy of the remnant and it slows further 
with $R_{SNR}\propto t^{2/7}$ (e.g. \cite{shull89}). For individual 
supernova remnants, the tracks may differ by less than $\pm 15\%$ 
for a factor of three variation in supernova kinetic energies 
(\cite{woosley95}). Also overplotted as dotted lines are the 
tracks of neutron stars born at the origin of the supernova 
explosion with varying velocities, showing the times required 
for them to catch up with the outer (radio) shell of the 
expanding remnant. 

We see from Figure $3$, that in spite of the uncertainties in the 
remnant ages and distances, all of the supernova shell remnants
associated with SGRs and AXPs seem to reside in the denser 
($n > 0.1$ cm$^{-3}$) phases of the ISM, where $<20 \%$ of the 
neutron star producing supernovae occur. Although the ages of 
the the SNRs are highly uncertain, information on the progenitor 
ISM density can be extracted from the SNRs because densities inferred 
from the Sedov and Radiative phase expansion formulae depend much more 
strongly on the SNR radii than on their ages. As we discuss in more 
detail below, the inferred distribution of progenitor ISM densities is 
not consistent with an origin of SGRs and AXPs that results from a purely 
intrinsic property of the neutron stars, because such stars should 
predominantly be born in the hot, diffuse ($n \sim 10^{-3}$ cm$^{-3}$) 
phase of the ISM, where $>80\%$ of the neutron star and pulsar producing 
SN occur. The probability of finding 10 such SNRs out of 12 possible 
occurring in the denser ($n > 0.1$ cm$^{-3}$) phases of the ISM, when 
at most $\sim 20\%$ are expected is only $\sim [12!/(10!2!)]{(0.2)}^{10}
{(0.8)}^{2} \sim 4\times 10^{-6}$. Even if we assume the maximum $16\%$ 
probability for chance associations with SNR for all SGRs and AXPs {\it 
and} assume that all chance associations will be in denser ISM, the 
combined probability of any single, typical SNR being in the denser 
ISM is 0.33, equalling $0.2$ naturally occurring plus $0.8\times 0.16$ 
occurring by chance. Thus the probability of finding 10 out of 12 possible 
occurring in the denser ISM is still only $\sim [12!/(10!2!)]{(0.33)}^{10}
{(0.67)}^{2}$ or $\sim 4\times10^{-4}$. 

There are two obvious ways that the distribution of these SNRs might 
be made consistent with that of typical neutron stars and pulsars, i.e. 
$80\%$ residing in the hot diffuse ISM and $20\%$ in the dense ISM. 
One way would be if the SNR distances were systematically underestimated 
by roughly a factor of 3, but as we saw in Figure 1 and 2, that is not
consistent with Galactic structure and would place roughly half of
the SNRs well outside the Galaxy. Alternatively, exceedingly weak 
($E<<10^{51}$ ergs) explosive energies for the SGR and AXP supernovae 
could in principle explain the unusually compact remnants associated 
with SGRs and AXPs, but this is inconsistent with dynamical requirements 
of supernova explosions (e.g. Woosley \& Weaver 1995), and with the 
direct observations of SGRs and AXPs associated SNRs (e.g. \cite{vancura92}; 
\cite{rho97}; \cite{gotthelf97}; \cite{long91}). Moreover, there is 
direct evidence that at least 8 out of the 10 SGRs and AXPs associated 
SNRs are indeed in dense environments. OH maser emission, attributed 
to SNR shock interactions with molecular clouds, has been detected 
(Frail et al. 1996; Green et al. 1997) from molecular clouds thought 
to be associated with five of the SNRs (CTB 33, Kes 73, W28, G10.0--0.3, 
and G346.6--0.2). CTB 109 (e.g. Huang \& Thaddeus 1985; Tatematsu et 
al. 1987), G287.8--0.5 (\cite{jones73}), and N49 (\cite{hughes84}) also 
show evidence for molecular clouds associations. These associations 
with molecular clouds clearly show that the supernova remnant shells 
associated with SGRs and AXPs are expanding in high density 
environments. 

\section{Comparison of SGRs/AXPs to Young Radio Pulsars}
\label{radiopulsars}

We can compare the observed properties of the SNRs associated 
with SGRs and AXPs with the remnants associated with another 
population of young neutron stars -- the young radio pulsars 
listed in the catalogs of Taylor, Manchester, \& Lyne (1993) and 
Taylor et al. (1995). We consider only the youngest pulsars with 
timing ages less than $\sim 30$ kyr, so as to be comparable to the 
age range in the SGR/AXP sample. The MDR timing ages of radio pulsars, 
unlike those for the SGRs and AXPs, are generally though to be good 
measures of their true ages (e.g. \cite{cordes98}), and we assume 
that the remnant ages are consistent with the MDR timing ages of 
their associated pulsars. A notable exception is PSR J1801--2451, 
which has an estimated age much greater than its MDR timing age 
(\cite{gaensler00}) -- possibly indicating the presence of non-MDR 
spin-down torques in this pulsar. For the pulsar distances, we 
use the distances derived (\cite{taylor93}; \cite{taylor95}) from 
the measured dispersion and the Taylor \& Cordes (1993) model of 
the free electron distribution in the Galaxy.  

The supernova remnant shells of many of the young pulsars in Table 
$3$ have not been detected in large scale radio surveys of the Galactic 
Plane (see e.g. \cite{whiteoak96}; \cite{reich84}; \cite{duncan97}), 
or in deep radio observations of the fields surrounding young pulsars 
(e.g. \cite{braun89}; \cite{frail94a}). For these pulsars, we found no 
likely associated SNRs in the Green (2000) catalog within a search radius 
corresponding to a transverse velocity of as much as $2000$ km s$^{-1}$, 
assuming the estimated distances and timing ages for these sources listed 
in Table $1$. Since the Green (2000) SNR catalog is more or less complete 
(\cite{whiteoak96}) down to a limiting surface brightness, the missing 
SNRs for these objects have probably expanded such that their surface 
brightnesses have faded below the limiting surface brightness of the 
radio surveys of the Galactic plane. We therefore assume that the radio 
shells of the remnants associated with these pulsars (except for the 
Crab, for which we use a lower limit of $17$ pc for the remnant 
radius; \cite{frail94a}) have expanded beyond this detectability 
threshold, and following Braun, Goss, \& Lyne (1989) we assign lower 
limits of $30$ pc to the undetected remnants corresponding to these 
pulsars. The observed and derived parameters for the young radio 
pulsars are listed in Table $3$, with references to the timing data, 
distances, and SNR shell radii. A plot of the supernova remnant shell 
radii versus age for the radio pulsars is shown in Figure $4$.

We clearly see from Figure $4$ that most of the young pulsars appear 
to have been born in the hot, diffuse phase of the ISM, as expected 
from other observations of their O and B star progenitors and their 
environments discussed in the preceding section. Only $\sim 5$ of the 
$16$ young pulsars have SNRs which may be expanding in the denser 
phases ($n>0.1$ cm$^{-3}$) of the ISM, which is consistent with the 
expected fraction of $\sim 20\%$ or less despite the small number 
statistics. These results are summarized in Table $4$.

\section{Discussion}  
\label{discussion}

\subsection{The Significance of the Environmental Evidence}
\label{significance}

As seen from Figures $3$ and $4$, the SGRs and AXPs appear to form 
in denser regions of the interstellar medium than the sample of young 
radio pulsars. The significance of the apparent disparity between 
the SGR/AXP and radio pulsar environments can be evaluated by using 
``survival analysis'' (\cite{miller81}) methods, which are 
statistical techniques incorporating ``censored'' data (e.g. upper 
limits) into data analysis. We first convert the age and SNR radii 
values of Tables 1 and 3 to ambient ISM densities $n$ using the standard 
formulae in Shull (1983) and using values for each age and SNR radius 
midway between the range of values listed in the tables. A total 
supernova kinetic energy of $10^{51}$ ergs was assumed for both 
the SGR/AXPs and radio pulsars. The lower limits on the radii 
of the undiscovered SNRs in Table 3 therefore became upper limits 
on the ambient densities for these SNRs. 

The resulting distributions of $n$ for the SGR/AXPs and radio pulsars 
were then tested for consistency with a single parent sample using 
the statistical analysis package ASURV Rev 1.2 (\cite{lavalley92}), 
which implements the methods presented in Feigelson \& Nelson (1985). 
The two-sample univariate nonparametric tests used to compare the 
distributions consisted of two versions of the Gehan's Generalized 
Wilcoxon test (permutation and hypergeometric variances), the Logrank 
test, and the Peto \& Peto and Peto \& Prentice Generalized Wilcoxon 
tests (see \cite{lavalley92} and references therein; \cite{miller81}; 
\cite{feigelson85}). These methods differ in their assumptions regarding 
the censoring process and in the ways which they weight the data, but 
we found that they gave results which were consistent to within two 
orders of magnitude. We use the statistical tests to compute the 
probability $P_{same}$ that the distribution of $n$ for the SGRs 
and AXPs is consistent with the same distribution for the young 
radio pulsars. For the full sample of SNR associations in Table $1$ 
we calculate $P_{same}=9\times 10^{-5}-6\times 10^{-4}$. If we exclude 
from the sample the objects with tentative SNR associations -- SGR 
1801--23 and AXPs 1709--40 and 1048--5937 -- we find $P_{same}=1\times 
10^{-4}-8\times 10^{-4}$. Finally, if we change the maximum possible 
remnant age from $30$ kyr to $20$ kyr (\cite{braun89}), we obtain 
$P_{same}=5\times 10^{-4}-3\times 10^{-3}$ for the full set of SNR 
associations in Table $1$, and $P_{same}=8\times 10^{-4}-5\times 
10^{-3}$ if we exclude the tentative SNR associations. The data 
therefore supports the conclusion that the progenitors of SGRs 
and AXPs exploded in denser environments than the progenitors 
of radio pulsars.  
 
This conclusion can be alleviated only if one assumes that a large 
fraction of the previously claimed SNR associations with AXPs and 
SGRs are spurious {\it and} the remnants truly associated with 
these objects have large diameters and are currently undetected. 
As discussed in $\S 3$, however, it is unlikely that more than a 
couple of the associations in Table 1 are spurious, given the 
distribution of supernova remnants in the Galactic plane. At least 
five of the SNR associations in Table $1$ would have to be spurious 
before one could conclude that the evidence for dense SGR and AXP 
progenitor was insignificant (i.e. the probability $P_{same}>0.01$). 
The association of SGR 1900+14 with SNR G42.8+0.6 has recently been 
questioned after the recent discovery (\cite{lorimer00}) of a 40 
kyr old pulsar PSR J1907+0918 equally close $\sim 20'$ to the SNR, 
and the discovery (\cite{vrba2000}) of a compact cluster of massive 
stars at an estimated distance ranging from $12$ to $15$ kpc, and only 
$\sim 10''$ from the SGR 1900+14 line of sight. The relatively low H 
column depth from absorption in the SGR's x--ray spectrum (\cite{hurley99b}; 
\cite{woods99a}) clearly implies (e.g. Figure $2$) a distance closer 
to 5 kpc, so that the SGR and and SNR may not be related to the cluster. 
If SGR 1900+14 is associated with the compact cluster at the more 
extreme distance, Vrba et al.(2000) suggest an association with a 
possible compact ($<1$ pc diameter) SNR indicated by the local diffuse 
x--ray emission. This new SGR/SNR association would therefore imply a much 
greater SGR 1900+14 progenitor density than the G42.8+0.6 association, 
increasing the evidence for dense progenitor environments for SGRs. 
A similar compact ($<1.0$ pc diameter) cluster of massive stars has 
also been observed (\cite{fuchs99}) to be coincident with SGR 1806--20 
and G10.0--0.3. In addition, AXP 1048--5937 and its associated SNR 
G287.8--0.5 are associated with the Carina star-forming region 
(\cite{the71}; \cite{seward82}). In these cases the associations 
with the clustered star-formation regions are quite consistent with 
the SGR/SNR associations, since the distances of the star-forming 
regions and the SNRs are comparable.

\subsection{SGRs and AXPs as Magnetars}
\label{magnetars}

The unusual properties of SGRs and AXPs may be due to the fact 
that they have unusually strong magnetic fields, as in the magnetar 
model (\cite{duncan92}; \cite{thompson93}; \cite{thompson95}; 
\cite{thompson96}). In the context of this model, it has been 
suggested that the association of SGRs and AXPs with dense 
progenitor environments may be explained by a selection effect 
in the following manner. Magnetars are thought to form from 
progenitor stars with high angular momenta, such that the 
superstrong ($10^{14}-10^{15}$ G) magnetic field can be 
generated in the protoneutron star by dynamo action just 
after core collapse (\cite{duncan92}). Since the total stellar 
angular momentum $J\propto MRv_{R}$ (assuming rigid rotation, where 
$v_{R}$ is the observed rotational velocity), one might expect the 
largest and most massive stars to be the likely progenitors of 
magnetars. Furthermore, since the main sequence lifetimes of stars 
decrease with increasing mass, the more massive stars would be the 
first to supernovae -- therefore exploding into more dense surroundings 
before the parent molecular cloud has been cleared by the successive 
supernovae of less massive stars. 

The two basic assumptions of this scenario do not appear to be 
consistent with observations. First, the observed stellar rotation 
velocities (e.g. \cite{drilling00}) show that the angular velocity 
$\Omega$ increases with stellar mass only up to about 2 M$_{\odot}$ 
and then {\it decreases} with increasing mass, dropping to only 
$20\%$ of its maximum value for stellar masses $>25M_{\odot}$. 
Since the stellar evolution timescale even in massive stars is 
much longer than the convective timescale (e.g. \cite{kippenhahn91}), 
the rotation of the star is effectively rigid and one would therefore 
expect the pre-collapse cores of the most massive stars to have 
less angular momenta than the cores of less massive supernova 
progenitors. This is supported by detailed calculations of the 
stellar evolution of rotating massive stars up to the time of 
core collapse (\cite{heger00}), which indicate that the final 
angular momentum of the core region is less for a massive star 
than for a less massive star with the same initial rotation rate. 
This is due to the greatly increased mass losses of luminous massive 
stars, which shed angular momentum in their powerful stellar winds 
(e.g. \cite{heger00}). 

Second, the supernovae of massive stars would still not be expected 
to all explode in the denser medium, because star formation in massive 
($>$ 10$^5$ M$_{\odot}$) clouds is episodic through several generations 
(e.g. \cite{mckee97}). These successive generations commence roughly 
every $\sim$4 to 5 Myr, over a period of $\sim$20 to 30 Myr, each 
producing roughly 10$^3$ O and B star supernova progenitors. Each 
subsequent generation of star formation commences shortly after the 
onset of supernova explosions from the previous generation, at the 
end of their main sequence lifetimes of $>$3 Myr. Therefore only the 
most massive stars in the first generation of star formation in a new 
star forming region would be expected to explode in denser environments, 
and the subsequent generations of massive stars should be born in the 
same low density superbubble environment as the population as a whole. 
Thus even if magnetars were produced by the most massive stars, we 
still would not expect more than a small fraction of them to explode 
in the denser phases of the ISM. 

\subsection{Propeller-based Models for SGRs \& AXPs}
\label{accretion}

The rapid spin-down rates, ages, clustered spin periods, and 
x--ray luminosities of AXPs and/or SGRs can all be explained by 
models involving the propeller effect as the dominant spin-down 
torque (\cite{vanparadijs95}; Alpar  2000; Chatterjee, Hernquist 
\& Narayan 2000; Chatterjee \& Hernquist 2000). In the context 
of these models, SGR bursts can be explained in terms of solid 
body accretion (\cite{hartwitt73}; \cite{tremaine86}) or crustal 
instabilities (\cite{blaes90}). Because there are upper limits on 
the masses of the possible binary companions of at least some of 
the AXPs (e.g. \cite{mer98}), the accreted material is probably 
ejecta from the neutron star's own supernova explosion 
(\cite{vanparadijs95}; \cite{corbet95}). In addition, since 
some of the SGRs and AXPs are located outside their apparent 
supernova remnants, the accreted material must form an accretion 
disk which can store the angular momentum of the accreted material 
and spin-down the neutron star on time scales of $1-10$ kyr. Upper 
limits on the disk emission from the nearby AXP 2259+58.6 (\cite{coe98}; 
\cite{hulleman00}) may rule out a standard hydrogen accretion disk 
around this AXP, but a disk may still exist given the uncertainties 
in the spectra of disks formed from metal-rich supernova ejecta and 
dust (e.g. \cite{perna00}). In addition, the noisy spin-down of SGR 
1806--20 (\cite{woods00}) and the $6.4$ keV emission feature 
(\cite{strohmayer00}) observed from an SGR 1900+14 burst are 
both consistent with accretion disks around these objects. There 
are two possible scenarios involving accretion from the ejecta of 
the supernova explosion which produces the neutron star: fallback 
disk accretion, ``pushback'' disk accretion, and accretion 
involving high velocity neutron stars.   

It was recently proposed that AXPs may be formed from neutron 
stars accreting material from ``fallback'' accretion disks 
(Chatterjee, Hernquist \& Narayan 2000; Chatterjee \& Hernquist 
2000). These disks may be formed from $\sim 0.001-0.1M_{\odot}$ 
(\cite{michel88}; \cite{lin91}) of inner ejecta material $<2$ hours 
after the initial core collapse in a type II supernova explosion 
(e.g. \cite{chevalier89}; \cite{woosley95}). Since a total accreted 
mass of only $10^{-6}M_{\odot}$ is required to explain the spin-down 
of SGRs and AXPs via the propeller mechanism (e.g. Alpar 2000; 
Chatterjee, Hernquist \& Narayan 2000; Chatterjee \& Hernquist 
2000), only a very small fraction ($\sim 10^{-5}$) of the fallback 
material must be accreted into a disk for this model to explain 
spin-down rates and ages of the SGRS and AXPs. Formation of such 
a fallback disk is limited to $<7$ days after the core collapse 
because of heating of the ejecta by $^{56}$Ni (\cite{chevalier89}). 
This is long before the remnant feels the external environment 
early in the Sedov phase, and therefore formation of an early 
fallback disk may not be compatible with the evidence for 
dense SGR and AXP progenitor environments. 

A model involving fallback disks which form later in the evolution 
of the SNR, however, could be consistent with the evidence of dense 
SGR and AXP environments. In particular, the expansion of slow-moving 
ejecta can be reversed by the Sedov phase reverse shock resulting 
from the interaction between the blastwave and the external medium 
(e.g. \cite{mckee74}; \cite{truelove99}), and the subsequent implosion 
could result in the formation of a ``pushback'' disk -- so named because 
the SNR ejecta is pushed-back onto the star because of the interaction 
with the dense environment surrounding the progenitor. Pushback disk 
formation would probably be most likely for: 1) neutron star progenitors 
which experienced a large amount of mass loss prior to supernova, and 2) 
progenitors in the dense ISM, which would provide the necessary pressure 
to confine the wind mass near the star. This is exemplified by the 
circumstellar environments surrounding the progenitor of SN 1987A, 
which is surrounded by $n\sim 10^{2}-10^{3}$ in wind material and 
HII gas (e.g. \cite{chevalier95}). The pushback process begins at 
the ``reversal time'' $t_{rev}\sim 0.5-1.0$ kyr (\cite{truelove99}), 
and the pushback mass is $\sim 0.4M_{\odot}$ given a total ejecta 
mass of $10 M_{\odot}$. As with the fallback disks discussed above, 
the formation of a pushback disk from only a small fraction of this 
matter would be required to explain the spin-down of SGRs and AXPs. 
Since this later fallback occurs after the majority of $^{56}$Ni and 
$^{56}$Co decays, the formation of the accretion disk around the neutron 
star would not be limited by radioactive heating. Such a model may be 
a plausible explanation for the dense SGR and AXP environments, and 
needs to be explored in more detail.  
  
The final propeller-based scenario for SGRs and AXPs involves high 
velocity neutron stars (HVNSs) capturing disk material from co-moving 
supernova ejecta, as first suggested Van Paradijs, Taam, and van den 
Heuvel (1995). Although the exact mechanism by which neutron stars are 
given substantial ``kick'' velocities at birth is not known, observations 
show that the kick velocities exceed $500$ km s$^{-1}$ in approximately 
$20\%$ of all neutron stars (\cite{cordes98}). In addition, this kick 
velocity appears to be independent of the dipole moment of the 
neutron star, as indicated by population studies of isolated radio 
pulsars (\cite{cordes98}; \cite{deshpande99}) and observations of 
extremely high velocity stars with canonical neutron star magnetic 
fields of $\sim 10^{12}$ G (e.g. PSR B2224$+$65: \cite{romani97}). 
As implied by the ratio $\theta_{\ast}/\theta_{SNR}$ listed in Table 
$1$, many of the SGR/AXP positions are significantly displaced from 
the apparent centers of their associated SNRs. These displacements 
imply that the SGR/AXPs may have systematically large transverse 
velocities, although there is considerable uncertainty in the actual 
velocities, mainly because of uncertainties in the ages of the 
associated remnants. In addition, the actual space velocities of 
the SGR/AXPs are larger by an unknown factor dependent on the 
viewing angle. It has been estimated (\cite{vanparadijs95}) that 
a $10^{-4}M_{\odot}$ accretion disk may be acquired by a high velocity 
neutron star as it moves through nearly co-moving supernova ejecta. 
However detailed calculations of the time-dependent accretion rate 
for the range of ISM densities, progenitor mass loss parameters, 
and neutron star magnetic fields, initial spin periods, and 
velocities are needed to properly constrain such a model. This 
is beyond the scope of the present paper, and will be left for 
future work.

Additional evidence in favor of the pushback disk and HVNS models 
for SGRs and AXPs may be provided by the observed number of these 
sources. For the pushback disk model, the number of SGRs and AXPs 
less than $t$ years old is given by $N=r_{b}f_{w}f_{env}t$, where 
$r_{b}$ is the Galactic neutron star birthrate, $f_{env}$ is the 
fraction of neutron star progenitors in the warm dense ISM, and 
$f_{w}$ is the fraction of neutron star progenitors which experience 
mass loss sufficient to form a pushback disk in dense ISM environments. 
Since the rate of mass loss from stellar winds is an increasing 
function of the initial main sequence mass, the fraction $f_{w}$ 
can be estimated by considering the {\it minimum} stellar mass which 
undergoes pre-supernova mass loss sufficient to decelerate its supernova 
ejecta at early times. To first order, this should occur when the total 
wind mass emitted during the progenitor's life is approximately equal 
to the supernova ejecta mass. From the solar metallicity stellar models 
of Schaller et al. (1992) -- which include mass loss -- this occurs for 
stars of initial main sequence masses greater than $M_{min}\sim 27
M_{\odot}$. Using $M_{min}$ and a Salpeter IMF with a maximum and 
minimum neutron star progenitor mass of $40 M_{\odot}$ and 
$8M_{\odot}$, respectively (\cite{woosley95}), we obtain $f_{w}\sim
0.1$. Assuming $f_{env}<0.2$ (as discussed in $\S 2$) and $r_{b}\sim 
1/40$ yr$^{-1}$ (\cite{van94}), we obtain $N<15$ SGRs and AXPs with 
ages less than $t=30$ kyr. This is similar to the observed number 
($12$) of these sources, which can be taken as evidence supporting the 
pushback disk model. A similar calculation is possible for the HVNS 
model. In this case, the expected number of SGRs and AXPs less than $t$ 
years old is given by $N=r_{b}f_{hv}f_{env}t$, where $f_{hv}$ of neutron 
stars with high space velocities and $r_{b}$ and $f_{env}$ are defined 
as before. Using the same values of $f_{env}$ and $r_{b}$, and assuming 
$f_{hv}\sim 0.2$ (velocities $>500$ km s$^{-1}$; \cite{cordes98}), yields 
$N\sim30$ expected SGR and AXP sources with ages less than $30$ kyr for 
the HVNS model. Again, these numbers are in the right ballpark for the 
observed numbers of SGRs and AXPs. 

\section{Summary} 
\label{summary}

We have shown that soft gamma--ray repeaters (SGRs) and anomalous 
x--ray pulsars (AXPs) are born in regions of the interstellar medium 
which are denser than the environments typical of young neutron stars. 
This suggests that the development of SGRs and AXPs may be related to 
their environments, and we examine the implications of this on magnetar 
and propeller-based models for SGRs and AXPs. The evidence of dense 
progenitor environments would be consistent with the magnetar model 
only if magnetars are born exclusively in dense environments, which does
not appear to be the case if magnetars form only from the most massive 
stars. Propeller-based models for SGRs and AXPs involving the formation 
of accretion disks from supernova ejecta appear to be consistent 
with the evidence for dense progenitor environments since these
environments may induce the formation of such disks. This may occur 
in two ways. {\it Pushback} disks may be formed from the infall of
the innermost ejecta, pushed back towards the neutron stars by prompt 
reverse shocks from the interactions of the expanding remnants with 
massive progenitor winds confined close to the stars by dense 
surrounding gas -- producing rapid deceleration of the expanding 
ejecta and strong prompt reverse shocks (Truelove \& McKee 1999). 
Fossil disks may also form around high velocity neutron stars accreting 
from nearly co-moving supernova ejecta, slowed by the strong prompt 
reverse shocks in such dense environments (van Paradijs et al. 1995).

\acknowledgements

We acknowledge helpful suggestions from anonymous referees 
which led to improvements in the paper. This research made 
extensive use of NASA's Astrophysics Data System Abstract 
Service. This work was performed while one of the authors 
(DM) held a National Research Council-GSFC Research Associateship. 
RER acknowledges support by NASA contract NAS5-30720, and REL 
support from the Astrophysical Theory Program.

\appendix

\section{Notes On Individual SGRs \& AXPs}

Here we discuss and reference the SNR distances, radii, and ages 
listed in Tables $1$ and $2$. The SNRs associated with all but three 
of the SGRs and AXPs have kinematic distances derived from associations 
with objects (e.g. HII regions or molecular clouds) having known 
distances. The distances of the other three SNRs -- G29.6+0.1, 
G346.6--0.2, and G42.8+0.6 -- are derived from consideration of 
Galactic spiral arms along the line of sight. The remnants Kes 
73, N49, W28, and CTB 109 have published age estimates based on 
measurements of shock velocities and/or x--ray temperatures, and 
for these remnants we adopt an age range corresponding to the 
minimum and maximum values quoted in the literature. The rest of 
the associated remnants have no published age estimates, other 
than those determined by assuming Sedov expansion and distances 
determined from $\Sigma-D$ relations, which are biased toward denser 
phases of the ISM. For these SNRs we adopt a broad range of possible 
ages. We take a lower age limit of $t_{low}=\rm{max}[t_{fe},t_{v}]$, 
where $t_{fe}$ and $t_{v}$ are the minimum ages derived by assuming 
free expansion of the SNR (at $10^{4}$ km s$^{-1}$) and a maximum 
SGR/AXP transverse velocity of $2000$ km s$^{-1}$, respectively. 
For an upper age limit, we take $30$ kyr, which is the maximum 
estimated age for a SNR/radio pulsar association in Table 3. Finally, 
the hydrogen column densities ($N_{H}$, listed in Table $2$) are 
determined from the best-fit spectral models for the x--ray spectra 
of the SGRs/AXPs.

{\it AXP 1841--045}: This AXP is coincident with the SNR G27.4+0.0 
(Kes 73), which is estimated from the x--ray temperature to be $0.5-2.5$ 
kyr old (\cite{helfand94}; \cite{gotthelf97}). The diameter of the 
radio remnant is $\sim 4'$ (\cite{kriss85}), and the distance from 
HI absorption is estimated to be $6.0-7.5$ kpc (\cite{san92}). 
A power law fit to the x--ray spectrum of the AXP yields a best-fit 
value of $N_{H}=(2.7-3.4)\times 10^{22}$ cm$^{-2}$ (\cite{gotthelf97}),
which is consistent with such a distance (see Figure 2), and with
the $N_{H}$ of $\sim2\times 10^{22}$ cm$^{-2}$ found (e.g. 
\cite{vasisht97}) of the x-ray spectrum of the SNR.

{\it SGR 0526--66}: The error box of the SGR lies within the supernova 
remnant N49 (\cite{evans80}; \cite{cline82}), located in the LMC at a 
distance of $49-55$ kpc (\cite{feast91}; \cite{capa90}) and interacting 
with a molecular cloud (\cite{hughes89}). The diameter of N49 is $\sim 
1'$, and its estimated age is $5-16$ kyr (\cite{vancura92}; \cite{shull83}).

{\it AXP 2259$+$58.6}: The AXP lies within the $\sim 30'$ diameter 
(\cite{hughes84}) SNR CTB 109 (G109.2--1.0). The distance to the 
remnant, as determined by spectroscopy of stars in nearby HII regions, 
is in the range $3.6-5.5$ kpc (\cite{rho97}). The estimated remnant age 
is $3.0-17.0$ kyr (\cite{parmar98}; \cite{hughes81}), and there is 
evidence for interaction between CTB 109 and surrounding molecular clouds 
(e.g. \cite{tatematsu87}). The best-fit model to the x--ray spectrum 
of the AXP yields $N_{H}=(0.9-1.0)\times 10^{22}$ cm$^{-2}$ 
(\cite{parmar98}), which agrees with the range of $N_{H}=(0.9-1.1)
\times 10^{22}$ cm$^{-2}$ found for the x-ray spectra from various 
parts of the remnant (\cite{rho97}).

{\it AXP 1845--0258}: This AXP is lies within the $5'$ diameter 
remnant G29.6+0.1 (\cite{gaensler99}). There are no kinematic 
distances, and the distance estimated from Galactic structure 
(see Figure $1$) is $9.0-13.5$ kpc. This is consistent (see Figure 
$2$) with the AXP x--ray absorption of $N_{H}=(9\pm 1)\times 10^{22}$ 
cm$^{-2}$ (\cite{torii98}). There are no reliable SNR age estimates, 
so we adopt an SNR age range of $0.6-30.0$ kyr by taking the limits 
of free expansion and the maximum detected SNR age from the radio 
pulsar sample. The chance association probability between the AXP 
and the SNR is estimated to be $1.6\times 10^{-3}$ (\cite{gaensler99}). 

{\it SGR 1627--41}: This SGR was first associated with the SNR 
G337.0--0.1 by Hurley et al. (1999c). The distance to this remnant 
is estimated to be $11.0\pm 0.3$ kpc (\cite{corbel99}), based 
on an association with a giant molecular cloud within the star-forming 
region CTB 33. The angular diameter of SNR G337.0--0.1 is $\sim 1.5'$ 
(\cite{corbel99}). The SNR age is unknown, so we adopt an age of 
$2.6-30.0$ kyr. The probability that the SGR/SNR association is spurious 
was estimated to be $\sim 5\%$ (\cite{smith99}). A marginal detection 
of a $6.4$ s pulsation from the SGR was reported (\cite{woods99b}) but 
not confirmed by subsequent observations (\cite{hurley2000b}). The x--ray 
spectrum of the SGR is equally well-fit by power law, blackbody, and 
thermal bremsstrahlung functions (\cite{hurley2000b}), and the mean 
$N_{H}$ for the power law spectral fit is $(7.4\pm 0.6)\times 10^{22}$ 
cm$^{-2}$ (\cite{woods99b}; \cite{hurley2000b}), which is consistent 
with the SNR distance (see Figure 2).

{\it SGR 1801--23}: The most recently discovered SGR, this source has 
a long and thin error box (\cite{cline00}) which passes through the 
center of SNR W28 (G6.4--0.1). W28 is associated with OH masers in
a molecular cloud whose distance rules out its previously suggested
association with PSR B1758--23 (\cite{claussen97}). W28 has a radio 
diameter of $\sim 40'$ (\cite{andrews83}), an estimated age of $2.4-30.0$ 
kyr (\cite{long91}), and is located at a distance of $1.2-3.0$ kpc 
(\cite{clark76}; \cite{goudis76}). The low energy absorption 
in the x--ray spectrum of W28 is $N_{H}=(0.5\pm 0.1)\times 10^{22}$ 
cm$^{-2}$ (\cite{rho98}), but there is no $N_{H}$ estimate for the SGR 
since its burst spectra were were not measured to low enough energy to 
determine a value, and the persistent low energy x-ray counterpart of
the source has not yet been identified. 

{\it AXP 1709--40}: This AXP lies $\sim 8.5'$ form the center of
the $\sim 10'$ diameter (\cite{whiteoak96}) remnant G346.6--0.2. 
The chance association probability between the AXP and this remnant 
is $\sim 0.1$, from the method of $\S 3$. There are no published age 
estimates for the SNR, so we derive and age range of $3.6-30.0$ kyr 
from the transverse velocity and maximum SNR age limits. There is no 
kinematic distance estimate to the SNR, so we adopt a distance of 
$3-5$ kpc, which agrees with Galactic structure arguments placing 
it in or near spiral arms. This distance is also consistent (Figure
$2$) with the AXP x--ray absorption of $N_{H}=(1.81\pm 0.07)\times 
10^{22}$ cm$^{-2}$ (\cite{sug97}).

{\it SGR 1806--20}: This SGR lies within the $\sim 7.5'$ diameter 
(\cite{kulkarni94}) remnant G10.0--0.3, which is located at a distance 
of $14.5\pm 1.4$ kpc as determined from CO line observations of 
maser-associated molecular clouds (\cite{corbel97}). This distance is 
consistent (Figure $2$) with the large column density $N_{H}=(6.0\pm 0.2)
\times 10^{22}$ cm$^{-2}$ found (\cite{sonobe94}) for the x--ray spectrum 
of the SGR. The chance SGR/SNR association probability has been 
estimated at $\sim 5\times 10^{-3}$ (\cite{kulkarni93}). The age is 
unknown, so we adopt an age of $3.5-30.0$ kyr. This SGR and SNR may be 
associated with a compact cluster of stars (\cite{fuchs99}). 

{\it SGR 1900$+$14}: This SGR has been associated (e.g. \cite{hurley99a})
with the $24'$ diameter remnant SNR G42.8+0.6 (\cite{furst87}) from whose
center it is displaced by $17'$. No kinematic distances to the SNR are 
available, so we adopt a distance of $3-9$ kpc from association with a 
Galactic spiral arm (Figure 1). This distance is also consistent (see 
Figure $2$) with the mean value of the SGR x--ray absorption 
for the best-fit spectral model: $N_{H}=(2.1\pm 0.2)\times 10^{22}$ 
cm$^{-2}$ (\cite{woods99a}). The age of the remnant is unknown, 
so we adopt an age range of $9.6-30.0$ kyr. The chance probability 
of the SGR/SNR association by the method of $\S 3$ is $\sim 0.1$. 
This SGR has also been associated (\cite{vrba2000}) with a compact star 
cluster at a distant of anywhere from $5$ to $15$ kpc, although the 
relatively low value of $N_{H}$ from the absorption in the SGR x--ray 
spectrum (\cite{hurley99b}; \cite{woods99a}) favors the near end of the 
range. 

{\it AXP 1048--5937}: This AXP is situated near the $\sim 25'$ 
diameter remnant G287.8--0.5 (\cite{jones73}) associated with 
the Carina star-forming region at a distance of $2.5-2.8$ kpc 
(\cite{the71}; \cite{seward82}). The geometrical chance association 
probability ($\S 3$) between the AXP and the remnant is $\sim 0.16$. 
There are no age estimates of the SNR and therefore we derive an age 
of $9.8-30.0$ kyr from the transverse velocity limit and the maximum 
detectable SNR age. The best-fit spectral model of the AXP emission
yields the low value of $N_{H}=(0.45\pm 0.10)\times 10^{22}$ cm$^{-2}$ 
(\cite{oos98}), which is consistent (e.g. Figure $2$) with the SNR/Carina 
distance.

{\it AXP 0720--3125}: Despite its low x--ray luminosity ($\sim 
5\times 10^{31}$ ergs s$^{-1}$, \cite{haberl97}), this $8.39$ s pulsar 
has been included in the AXP sample due to its spin period, lack of 
an optical counterpart (\cite{motch98}), and fast spin-down rate ($\sim 3
\times 10^{-12}$, \cite{haberl97}). Although there is no adjacent SNR 
in the catalog of Green (2000), the low x-ray absorption indicates 
a distance of only $0.10\pm 0.02$ kpc (\cite{haberl97}) to the source, 
which makes the detection of its associated SNR unlikely (see $\S 3$)
if its age is as great ($\sim 40$ kyr) as that indicated by its spin
down rate.
 
{\it AXP 0142$+$615}: This apparently old (possibly $\sim 60$ kyr) 
and well-studied (e.g. \cite{white96}; \cite{israel94} and references 
therein) AXP has no remnant within $\sim 1^{\circ}$ of it in the Green 
(2000) catalog. But there is evidence (\cite{white96}) that this source 
is in or behind a giant molecular cloud. Given the rapid evolution (e.g. 
\cite{truelove99}) and possible self-absorption (\cite{reynolds88}) of 
the radio emission of SNRs in dense environments, the associated SNR may 
have faded below detectability, and therefore we are unable to constrain 
the radius of the associated remnant. 
    
{}

\newpage

\begin{figure}
\plotone{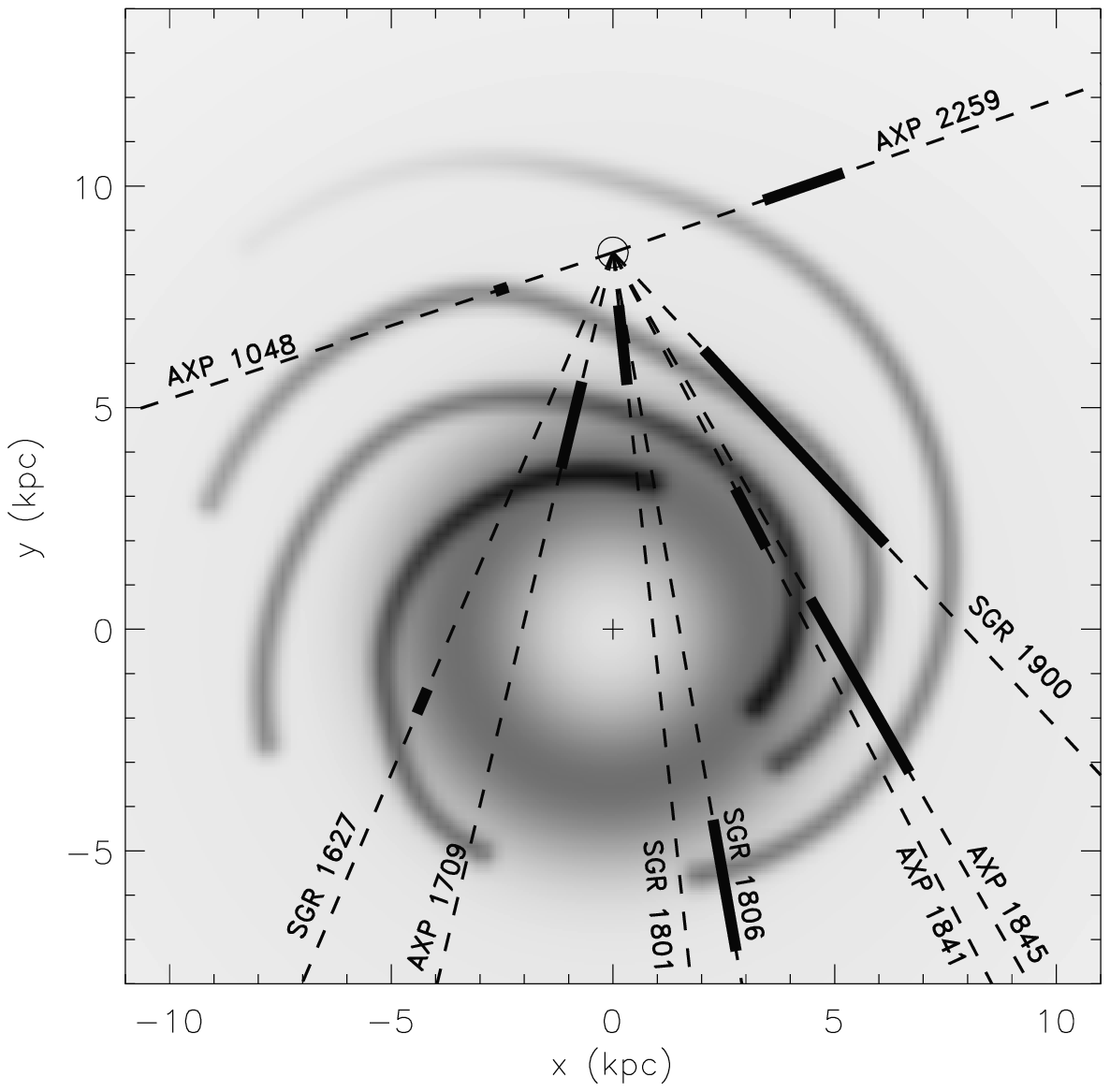}
\caption{~The positions of the soft gamma-ray repeaters (SGR) and anomalous 
x-ray pulsars (AXP) and their associated supernova shell remnants (SNR) 
projected onto a model (\cite{taylor93a}) of the Galactic Plane distribution 
of free electrons. The position of the Galactic Center and the Sun are 
marked by the plus (``+'') and circle (``$\circ$'') symbols, respectively, 
and the line of sight to each source is shown by a dashed line. The estimated 
distances to the sources, from Galactic kinematic arguments or association 
with molecular clouds or spiral arms, are shown by the heavy solid lines.}
\end{figure}

\begin{figure}
\plotone{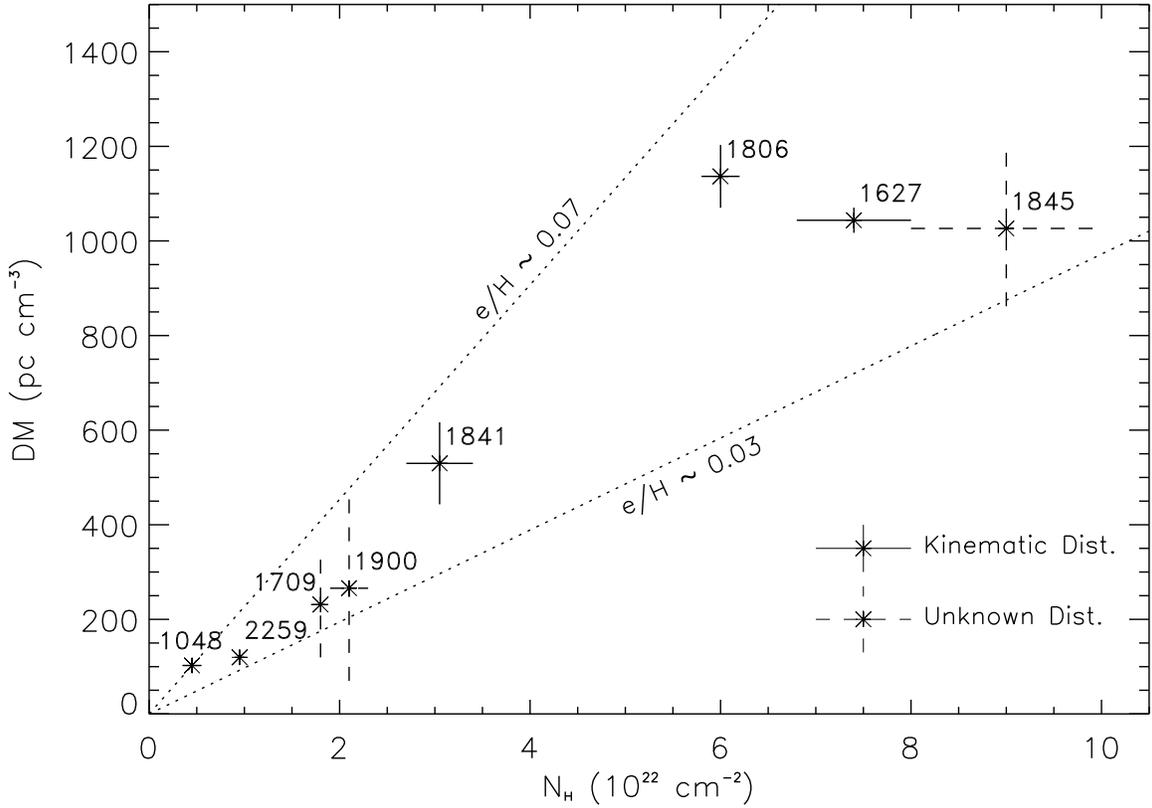}
\caption{~The inferred dispersion measure (\cite{taylor93a}) towards each 
of the SNRs, given the range of assumed distances, plotted against the 
observed $N_{H}$ values for the x-ray absorption of their associated SGRs 
and AXPs. As shown by the dotted lines, the data are consistent with the
typical range of Galactic electron to hydrogen ratios (e/H) (\cite{spitzer78}).
This supports the SGR/AXP--SNR associations by showing that the SNR distances 
are consistent with the SGR and AXP distances.}
\end{figure}

\begin{figure}
\centerline{\includegraphics[width=5.5in,keepaspectratio,angle=90]
{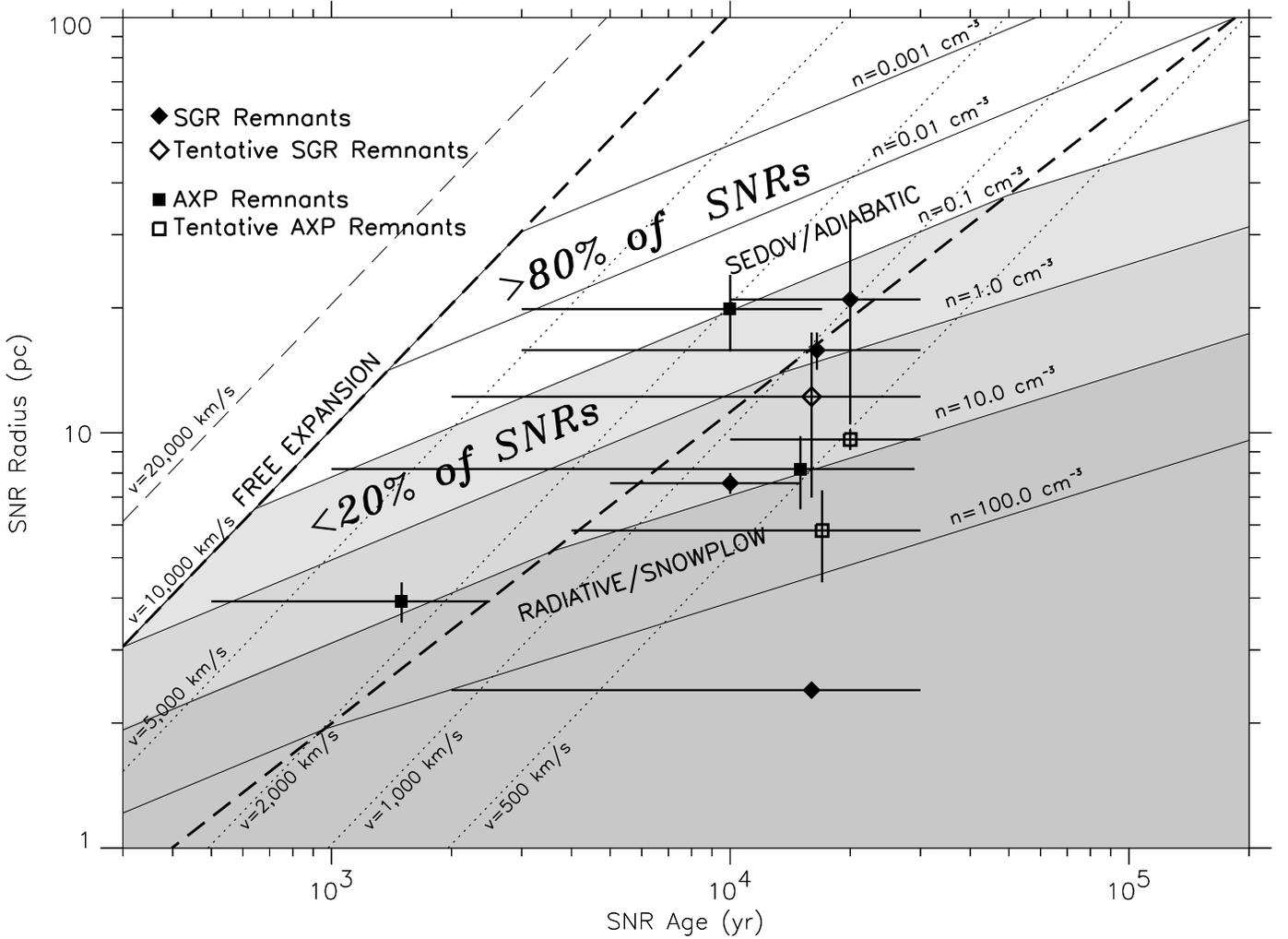}}
\caption{~The radius of the SGR and AXP supernova remnant shells as a 
function of their age. The solid lines denote SNR expansion trajectories 
in the free expansion, Sedov and radiative phases (separated by dashed 
lines), according to Shull, Fesen, \& Saken (1989), for an assumed
supernova ejecta energy of $10^{51}$ ergs in a wide range of ISM 
densities. The dotted lines denote the tracks of neutron stars born 
at the origin of the supernova explosion with varying space velocities. 
The data show that these objects are unusual in that they are all 
preferentially formed in the denser ($>0.1$ H cm$^{-3}$) phases of 
the interstellar medium (ISM), where $<$20\% of all neutron-forming 
supernovae occur, as determined from observations of OB associations 
and Galactic supernova remnants.}
\end{figure}

\begin{figure}
\centerline{\includegraphics[width=5.5in,keepaspectratio,angle=90]
{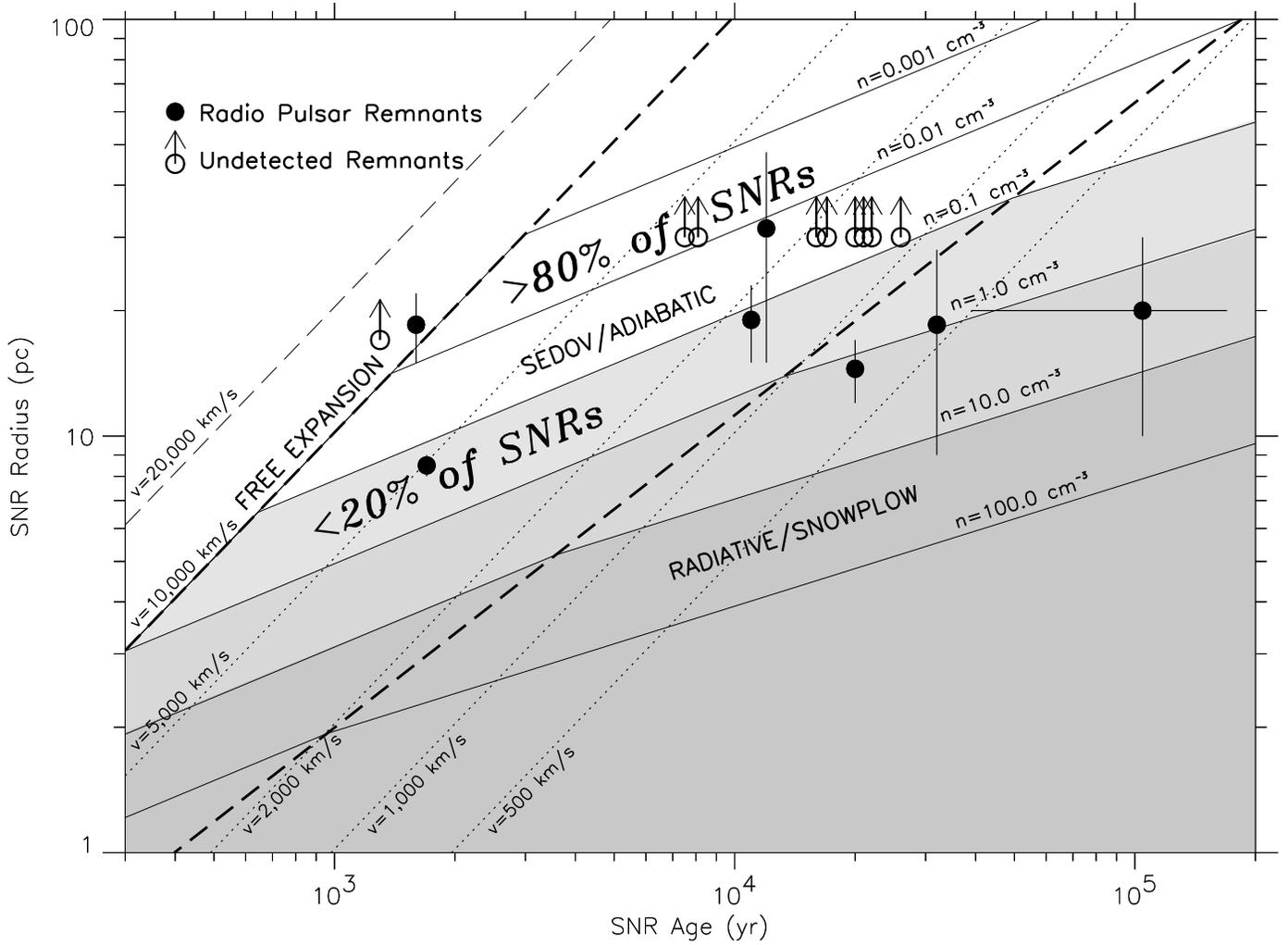}}
\caption{~Same as Figure $3$, except for young radio pulsars with 
timing ages less than $32$ kyr. The remnants for these pulsars reside 
primarily in the diffuse phase of the ISM, as expected from observations 
of OB associations and Galactic supernova remnants.}
\end{figure}

\newpage

\begin{planotable}{lccccr}
\tablewidth{0pt}
\tablecaption{The Supernova Remnants of SGRs and AXPs}
\tablehead{\colhead{Object} & \colhead{Period} & 
\colhead{SNR} & \colhead{Age\tablenotemark{a}} & 
\colhead{Rad.\tablenotemark{b}} & 
\colhead{${\theta_{\ast}\over \theta_{SNR}}$} \\ { } & 
(s) & { } & (kyr) & (pc) & { }}
\startdata
AXP 1841--045 & $11.8$ & Kes 73 & $1.5\pm1.0$ & $3.5-4.4$ & $0.1$ \nl
SGR 0526--66 & $8.0$ & N49 & $10\pm5$ & $7.1-8.0$ & $0.8$ \nl
AXP 2259$+$58.6 & $6.98$ & CTB 109 & $10\pm7$ & $16.-24.$ & $0.2$ \nl
AXP 1845--0258 & $6.97$ & G29.6$+$0.1 & $15\pm14$ & $6.5-9.8$ & $0.1$ \nl
SGR 1627--41 & $6.4$? & G337.0$-$0.1 & $16\pm14$ & $2.3-2.5$ & $2.3$ \nl
SGR 1801--23 & -- & W28\tablenotemark{c} & $16\pm14$ & $7.0-17.5$ & 
$0.1$ \nl
AXP 1709--40 & $11.0$ & G346.6--0.2\tablenotemark{c} & $17\pm13$ & 
$4.4-7.3$ & $1.7$ \nl
SGR 1806--20 & $7.47$ & G10.0--0.3 & $17\pm13$ & $14.-17.$ & $0.5$ \nl
SGR 1900$+$14 & $5.16$ & G42.8$+$0.6 & $20\pm10$ & $11.-31.$ & $1.4$ \nl
AXP 1048--5937 & $6.45$ & G287.8--0.5\tablenotemark{c} & $20\pm10$ & 
$9.1-10.2$ & $2.2$ \nl
AXP 0720--3125 & $8.39$ & --\tablenotemark{f} & $40$\tablenotemark{d} 
& -- & -- \nl
AXP 0142$+$615 & $8.69$ & --\tablenotemark{e} & $60$\tablenotemark{d} 
& -- & -- \cr 
\tablenotetext{a}{SNR age (see Appendix notes)}
\tablenotetext{b}{Radius of radio shell (see Appendix and distances in 
                  Table 2)}
\tablenotetext{c}{``Tentative'' remnant identification (see text)}
\tablenotetext{d}{MDR timing age since there is no identified SNR}
\tablenotetext{e}{In/behind molecular cloud (no identified remnant)}
\tablenotetext{f}{Too close to identify radio remnant}  
\end{planotable}

\newpage

\begin{planotable}{lccccccr}
\tablewidth{0pt}
\tablecaption{ Distance Measures of Galactic SGR and AXP and 
Associated SNRs\tablenotemark{a}}
\tablehead{\colhead{SNR} & \colhead{$l$} & \colhead{$b$} & 
\colhead{Dist.\tablenotemark{b}} & \colhead{DM\tablenotemark{c}} & 
\colhead{SGR/AXP} & \colhead{$N_{H}$} \\ { } & (deg.) & (deg.) & (kpc) 
& (pc cm$^{-3}$) & { } & ($10^{22}$ cm$^{-2}$)}
\startdata
Kes 73 & $27.386$ & $-0.006$ & $6.0-7.5$ & $440.-620.$ & AXP 1841--045 
& $2.7-3.4$\nl
CTB 109 & $109.093$ & $-0.993$ & $3.6-5.5$ & $100.-135.$ & AXP 2259$+
$58.6 & $0.9-1.0$\nl
G29.6$+$0.1 & $29.679$ & $-0.109$ & $9.0-13.5$\tablenotemark{d} & $860-
1200$ & AXP 1845--0258 & $8.0-10.0$\nl
G337.0$-$0.1 & $336.968$ & $-0.111$ & $10.7-11.3$ & $1020.-1070.$ & SGR 
1627--41 & $6.8-8.0$\nl
G346.6--0.2 & $346.482$ & $+0.035$ & $3.0-5.0$\tablenotemark{d} & 
$120.-340.$ & AXP 1709--40 & $1.7-1.9$\nl
G10.0--0.3 & $9.996$ & $-0.242$ & $13.-16.$ & $1070.-1200.$ & SGR 
1806--20 & $5.8-6.2$\nl
G42.8$+$0.6 & $43.021$ & $+0.766$ & $3.0-9.0$\tablenotemark{d} & 
$70-460.$ & SGR 1900$+$14 & $1.9-2.3$\nl
G287.8--0.5 & $288.257$ & $-0.518$ & $2.5-2.8$ & $90.-120.$ & AXP 
1048--5937 & $0.4-0.6$\cr
\tablenotetext{a}{See Appendix and references therein}
\tablenotetext{b}{Kinematic distance unless otherwise noted}
\tablenotetext{c}{Dispersion measures calculated from the model of Taylor 
\& Cordes (1993)}
\tablenotetext{d}{Adopted distance (see text)}
\end{planotable}

\newpage

\begin{planotable}{lccccrr}
\tablewidth{0pt}
\tablecaption{Young Radio Pulsars}
\tablehead{\colhead{Pulsar} & \colhead{Period} & 
\colhead{Age\tablenotemark{a}} & \colhead{Dist.\tablenotemark{b}} & 
\colhead{SNR} & \colhead{Rad.\tablenotemark{c}} 
& Reference \\ { } & (s) & (kyr) & (kpc) & { } & (pc) & { }}
\startdata
J0534$+$2200 & $0.033$ & $1.3$ & $1.5-2.5$ & $-$ & $>17$ & 1\nl
J1513--5908 & $0.151$ & $1.6$ & $3.5-5.3$ & MSH 15--52 & $15-22$ 
& 2\nl
J0540--6919 & $0.050$ & $1.7$ & $49.0-55.0$ & SNR 0540--693 & 
$8-9$ & 3\nl
J1614--5047 & $0.232$ & $7.5$ & $3.7-11.0$ & $-$ & $>30$ & 4\nl
J1617--5055 & $0.069$ & $8.1$ & $6.1-6.9$ & $-$ & $>30$ & 5\nl
J0835--4510 & $0.089$ & $11$ & $0.4-0.6$ & Vela XYZ & $15-23$ & 
11\nl 
J1341--6220 & $0.193$ & $12$ & $4.0-13.0$ & G308.8--0.1 & $15-48$ 
& 7\nl
J1801--2451 & $0.125$ & $39-170$ & $2.3-6.9$ & G5.4--1.2 & $10-30$ & 
8\nl
J1803--2137 & $0.134$ & $16$ & $2.9-4.9$ & $-$ & $>30$ & 8\nl
J1709--4428 & $0.102$ & $17$ & $1.4-2.7$ & $-$ & $>30$ & 2,10\nl
J1856$+$0113 & $0.267$ & $20$ & $2.7-3.9$ & W44 & $12-17$ & 
9\nl
J1048--5832 & $0.124$ & $20$ & $2.2-3.7$ & $-$ & $>30$ & $-$\nl
J1740--3015 & $0.607$ & $21$ & $2.5-4.1$ & $-$ & $>30$ & 6\nl
J1826--1334 & $0.101$ & $22$ & $3.1-5.2$ & $-$ & $>30$ & 6\nl
J1730--3350 & $0.139$ & $26$ & $3.2-5.3$ & $-$ & $>30$ & $-$\nl
J1646--4346 & $0.232$ & $32$ & $3.4-10.3$ & G341.2$+$0.9 & $9-28$ 
& 10\cr 
\tablenotetext{a}{Timing age (from \cite{taylor93}; \cite{taylor95}), 
except for J1801--2451 (\cite{gaensler00})}
\tablenotetext{b}{Distance from pulsar dispersion measure}
\tablenotetext{c}{Radius of SNR radio shell}
\tablerefs{(1) \cite{frail95}; (2) Caswell et al. 1981; (3) 
Manchester et al. 1993; (4) \cite{johnston95}; (5) \cite{kaspi98}; 
(6) Braun et al. 1989; (7) \cite{caswell92}; (8) \cite{frail94b}; (9) 
\cite{giacani97}; (10) \cite{frail94a}; (11) \cite{green84}}
\end{planotable}

\newpage

\begin{planotable}{lrr}
\tablewidth{300pt}
\tablecaption{Occurrences in the Warm and Hot Phases of the ISM}
\tablehead{\colhead{Source} & \colhead{WISM($\%$)\tablenotemark{a}} & 
\colhead{HISM($\%$)\tablenotemark{b}}}
\startdata
ExtraGal SNe & $<20$ & $>80$\nl
Galactic SNR & $10\pm 10$ & $90\pm 10$\nl
Young PSRs & $31\pm 14$ & $69\pm 21$\nl
AXP/SGRs & $83\pm 26$ & $17\pm 12$\cr 
\tablenotetext{a}{Percentage in the warm ISM ($n\geq 0.01$ cm$^{-3}$)}
\tablenotetext{b}{Percentage in the hot ISM ($n\sim 0.001$ cm$^{-3}$)}
\tablecomments{ExtraGal SNe -- from van Dyk et al. (1996) observations 
of 49  extragalactic Type II \& Ib/c SNe, corrected for detection 
threshold as discussed in text; Galactic SNR -- from Higdon \& 
Lingenfelter (1980) analysis of Clark \& Caswell (1976) catalog 
of Galactic SNRs; Young PSRs -- youngest ($<30$ kyr) pulsars, see 
Table 3 and Figure 4; AXP/SGRs -- see Table 1 and Figure 3.}
\end{planotable}

\end{document}